\documentclass[journal=nalefd,manuscript=communication%,layout=twocolumn
]{achemso}

\usepackage[T1]{fontenc} % Use modern font encodings
\usepackage{bm}% bold math
\usepackage{mathptmx} % Times

\usepackage{ulem}
\usepackage{xcolor}
\definecolor{darkgreen}{rgb}{0.,0.45,0.}
\definecolor{lightgreen}{rgb}{0.,0.76,0.76}
\definecolor{darkblue}{rgb}{0.1,0.2,0.46}
\definecolor{lightred}{rgb}{1.0,0.3,0.3}
\definecolor{darkred}{rgb}{.8,0.1,0.0}
\definecolor{lilac}{rgb}{0.6,0.2,0.7}

\newcommand\removed[1]{}

\renewcommand{\vec}[1]{\boldsymbol{\mathbf{#1}}}
\newcommand{\unitvec}[1]{\hat{\boldsymbol{\mathbf{#1}}}}
\newcommand{\FGT}[1][]{Fe$_{#1}$GeTe$_2$}

\newcommand{\SI}[0]{S.I.}

\author{João Sampaio}%
    \email{joao.sampaio@cnrs.fr}
\author{Antoine Pascaud}
\author{Edgar Quero}
\author{André Thiaville}
\affiliation{ Universit\'e Paris-Saclay, CNRS, Laboratoire de Physique des Solides, 91405 Orsay, France}
    
\author{Vincent Polewczyk}
\author{Alain Marty}
\author{Frédéric Bonell}
\affiliation{University Grenoble Alpes, CNRS, CEA, IRIG-Spintec, F-38000 Grenoble, France}

\author{Alexandra Mougin}
\affiliation{ Universit\'e Paris-Saclay, CNRS, Laboratoire de Physique des Solides, 91405 Orsay, France}

\title{Dzyaloshinskii-Moriya interaction in \FGT[5] epitaxial thin films}


\begin{document}

% The "tocentry" environment can be used to create an entry for the graphical table of contents. It is given here as some journals require that it is printed as part of the abstract page. It will be automatically moved as appropriate.
% \begin{tocentry}
%     \includegraphics[width=\textwidth]{Figures/TOC figure bls acs.pdf}
%     \added{Quantification of the Dzyaloshinskii-Moriya interaction in \FGT[5] epitaxial thin films versus thickness using Brillouin light scattering spectroscopy.}
% \end{tocentry}

% \date{\today}

\begin{abstract} %147 words
Van der Waals ferromagnets, such as \FGT[5], offer a promising platform for spintronic devices based on chiral magnetic textures, provided a significant Dzyaloshinskii-Moriya interaction (DMI) can be induced to stabilise the textures.
Here, we directly measure DMI in epitaxial \FGT[5] thin films using Brillouin light scattering spectroscopy and observe a consistent DMI ($D=0.04$~mJ/m$^2$) across various thicknesses. Its weak thickness dependence, combined with the nominally symmetric film interfaces, suggests a bulk origin. Although we do not determine the microscopic mechanism, our findings are compatible with \textit{ab initio} calculations linking DMI to partial ordering of Fe split sites. Additionally, we find a low magnetic dissipation ($\alpha<0.02$).
The observed DMI, which could be further enhanced by optimising the Fe site ordering, combined with low dissipation, makes \FGT[5] a strong candidate for exploring the dynamics of chiral magnetic textures in two-dimensional materials.
\end{abstract}

Ferromagnetic van der Waals (vdW) materials present an interesting new class of materials for the study of nanomagnetism. Their structure, made of 2D layers with relatively weak inter-layer bonding, can produce very thin films with flat and clean surfaces, either by exfoliation from macroscopic crystals or epitaxial growth. This structure also lends itself naturally to the fabrication of heterostructures with good interfacial quality with a large variety of other vdW materials (semiconductors, metals, insulators, topological insulators, ...) ---  ideal for studying interfacial magnetic effects that are the basis of spintronic devices. 
Among the growing list of vdW ferromagnets of interest to spintronics, the \FGT[x] family stands out due to their high Curie temperature \cite{May2019}, metallic character, and large magneto-transport effects.

Moreover, there have been reports of chiral magnetic structures in \FGT[x], opening the way to their use in the many proposed devices based on chiral textures such as skyrmions. Homo-chiral spin spirals were observed in exfoliated flakes of \FGT[3] \cite{Meijer2020} (thickness $t= 185$~nm), and Néel skyrmions were reported in \FGT[3] (70~nm)/[Co/Pd] \cite{Yang2020} and WTe$_2$/\FGT[3] (32~nm) \cite{Wu2020} heterostructures.  In other members of the \FGT[x] family, Néel skyrmions were observed in \FGT[5-x] \cite{Gao2022,Casas2023} (on 61 and 100 nm thick flakes), as well as in (Fe,Co)$_5$GeTe$_2$ (110 nm) \cite{Zhang2022}  or in the related Fe$_{3-x}$GaTe$_2$~\cite{Zhang2024,Li2024FeGaTe}. On the other hand, other groups working also on flakes of \FGT[5] observed only non-chiral Bloch bubbles \cite{Schmitt2022,Gopi2024}. 

The observation of homo-chiral textures suggests a significant Dzyaloshinskii-Moriya interaction (DMI). Indeed, a large DMI parameter ($D = 1.0$ mJ/m$^2$) was estimated in ref.~\citenum{Wu2020}. A nonzero DMI requires a symmetry-breaking mechanism along the out-of-plane direction. However, both \FGT[3] and \FGT[5] are presumed to have inversion-symmetric crystal structures incompatible with DMI\cite{Laref2020}. Nevertheless, several experimental studies suggest that the actual crystalline structure of \FGT[x] may lack inversion symmetry. Possible sources of symmetry breaking include Fe split-site ordering \cite{Ly2021}, Fe self-intercalation in the van der Waals gaps \cite{Silinskas2024}, ordered Te vacancies \cite{Gopi2024}, and asymmetric Fe vacancy distribution \cite{Chakraborty2022}. Additionally, deviations from the expected vdW stacking sequence (ABCABC in \FGT[5] and ABAB in \FGT[3]) could also break inversion symmetry.

DMI can also arise from asymmetric interfaces in thin magnetic layers. This mechanism has been well established in metallic systems, such as nanometric Co films in a AlOx/Co/Pt triplet film (up to $D=2$ mJ/m$^2$ \cite{Belmeguenai2015, Thiaville2012}). 
Although this interfacial mechanism could be expected to be  weaker in vdW materials due to their weak interlayer coupling, first-principle calculations by Park et al. \cite{Park2021} show that the interface of \FGT[3] with its natural oxide can induce a significant DMI ($\approx 2$ mJ/m$^2$ for a monolayer of \FGT[3]). 
Interfacial mechanisms, however, are only significant in ultra-thin systems, since their effects are progressively diluted with increasing thickness, while all the cited experimental observations involve relatively thick (30-200 nm) exfoliated flakes.

Characterizing the DMI solely from the observation of magnetic textures is challenging, as these are influenced by multiple interactions, including dipolar effects, magnetocrystalline anisotropy, and exchange. In some cases, dipolar interactions may even create seemingly chiral textures at the surface of thick samples \cite{Yang2020,Cheynis2009}. To determine the origin of chiral magnetic textures and disentangle interfacial and bulk contributions, a more direct measurement of the DMI is essential. In this letter, we investigate the static and dynamic magnetic properties of epitaxial \FGT[5] thin films of varying thickness using Brillouin light scattering (BLS) spectroscopy. By directly measuring the DMI, we unambiguously demonstrate the presence of a significant DMI of bulk origin. Furthermore, we show that the magnetic damping in these films is low for a metallic ferromagnet. These properties make \FGT[5] a promising material for the study of the propagation of chiral magnetic textures.

\begin{figure}[ht]
    \centering
    \includegraphics[width=0.48 \textwidth]{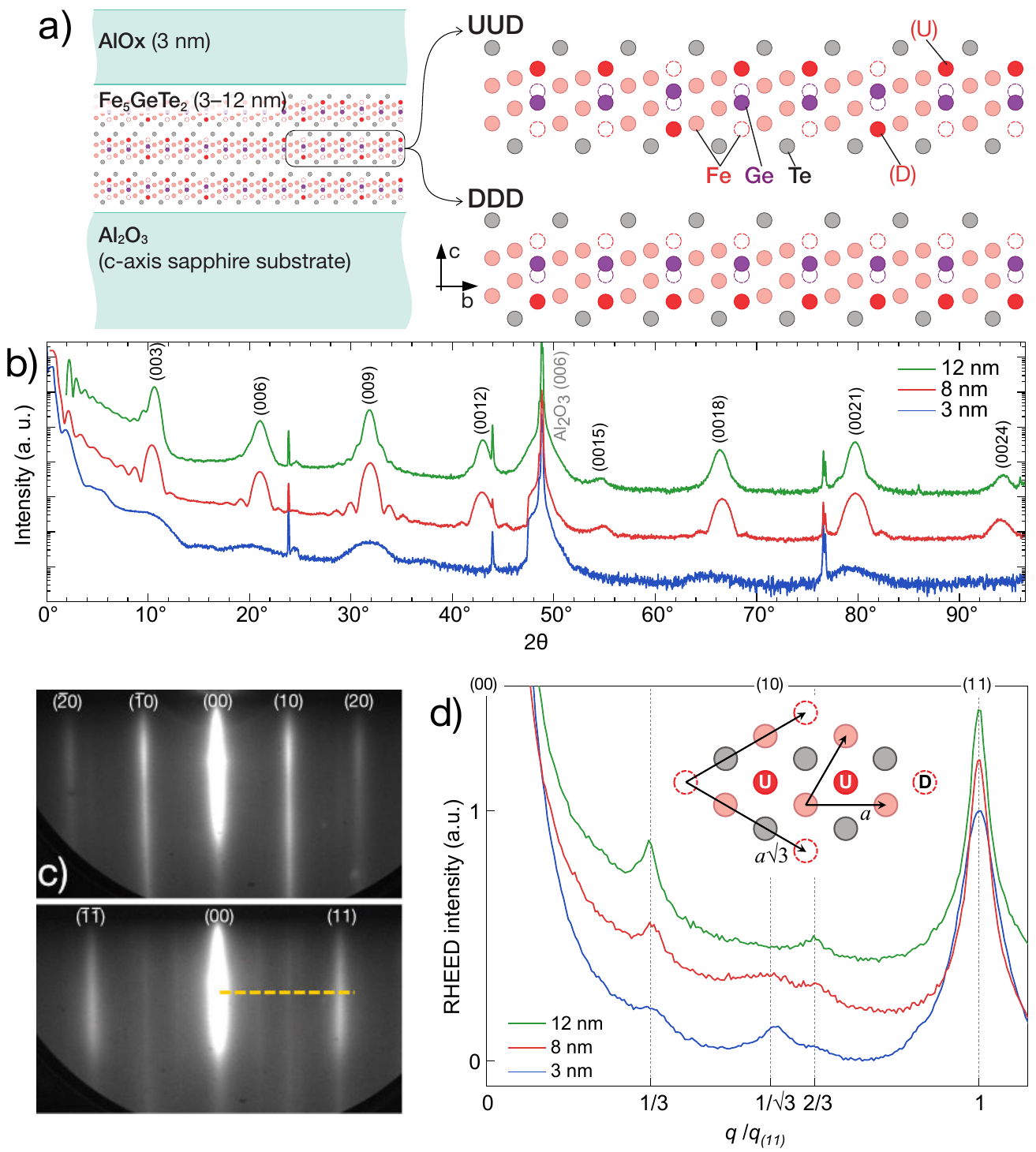}
    \caption{
    a) Sample structure and diagrams of ordered split sites \FGT[5] with broken centrosymmetry: full ordering (“DDD”) or 1/3 (“UUD”). Open circles represent empty sites. 
    b) XRD spectrum for the three films.
    c) RHEED along the (100) and (110) reciprocal directions, obtained after growth (3-nm thick film). 
    d) RHEED intensity profiles along the dashed line in c. The profiles are normalized to the (11) intensity and shifted vertically for clarity. 
    Inset: Top view schematic of the “UUD” $\sqrt{3}\times\sqrt{3}$ reconstruction.
    }
    \label{fig:Structure}
\end{figure}

\section{Results}

\subsection{Sample growth \& structure} 

\FGT[5] films ($t =$ 12, 8 and 3~nm) were epitaxially grown on Al$_2$O$_3$(0001)
at 350°C by coevaporation of elemental Fe, Ge and Te, with a Fe:Ge flux ratio of 5 and with a slightly overstoichiometric Te flux, and then annealed $\sim 500$°C for 10 min. 
The 12-nm-thick film was characterised in depth in Ribeiro et al. \cite{Ribeiro2022}, and all films were grown following the protocol reported in that work. The films were capped by 3~nm of Al deposited at room temperature, then naturally oxidized in air. This capping layer prevents the oxidation of \FGT[5] and provides similar top and bottom interfaces (Fig.~\ref{fig:Structure}a). 

The X-ray diffraction (XRD) specular scans (Fig.\ref{fig:Structure}b) show single-phase \FGT[5] with a constant $c$-axis lattice parameter (0.98 nm per monolayer) across all thicknesses (see \SI), consistent with the good crystalline quality previously observed for the 12-nm film by XRD and scanning transmission electron microscopy (STEM)~\cite{Ribeiro2022}.
The RHEED patterns of \FGT[5] (8 nm) in the (100) and (110) reciprocal directions, measured after growth, are shown in Fig.~\ref{fig:Structure}c. Similar anisotropic diffraction patterns are observed at all thicknesses and demonstrate the good epitaxy on sapphire. However, faint (10) diffraction rods are visible in the (110) direction (peak at $q/q(11)=1/\sqrt{3}$ in Fig.~\ref{fig:Structure}d). They denote the presence of a small proportion of domains that are rotated by 30° from the main crystal orientation. This feature is more apparent for the thinnest layer and disappears for the 12 nm film, indicating an improvement of the crystal in-plane orientation quality with increasing  thickness. 

In addition, the RHEED patterns display a ($3 \times 1$) reconstruction characteristic of a ($\sqrt{3} \times \sqrt{3}$) R30° superstructure (peaks at q/q(11) = 1/3 and 2/3 in Fig. \ref{fig:Structure}d). This superstructure is observed in all three films, and is more apparent at larger thicknesses. The nature of this feature will be discussed later.

\begin{figure}[ht]
    \includegraphics[width=0.45 \textwidth]{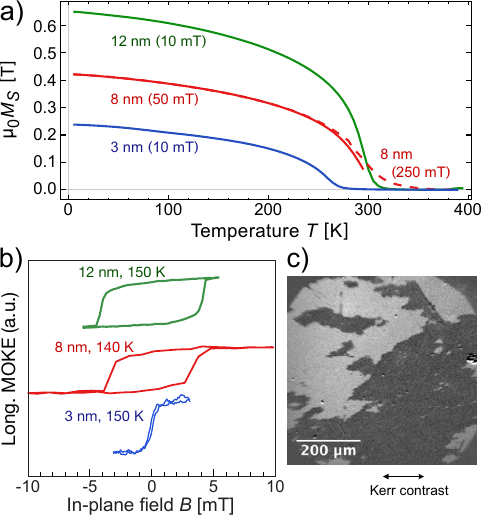}
    \caption{
    a) $M_s(T)$ of the three films under in-plane field (SQUID measurement).
    b) LMOKE hysteresis loops of the three films close to 150~K.
    c) In-plane magnetic domains in the 12-nm-thick sample imaged by LMOKE (the contrast orientation is left/right). }
    \label{fig:MagCaract}
\end{figure}

\subsection{Magnetic properties} 

The spontaneous magnetization ($M_s$) was measured by SQUID magnetometry versus temperature (Fig.~\ref{fig:MagCaract}). Thinner films showed lower $M_s$, with the 3-nm film showing about one-third of the  $M_s$  of the 12-nm film at 150 K (see Table~\ref{tab:parameters}). The Curie temperature ($T_C$) of the thicker films (297–299 K) is similar to bulk values (270–310 K \cite{May2019}), whereas the 3-nm film had a lower $T_C$ (258~K). The evolution of $M_s$ and $T_c$ indicates that a thickness of a few layers is needed to fully develop the bulk magnetic properties of the films. The $M_s(T)$ curves showed a monotonic trend, unlike previous reports of various transitions in $M_s(T)$ \cite{Alahmed2021,Schmitt2022,Ly2021,Lv2022}. However, those studies focused on virgin films, whereas our samples had been previously magnetized. These transitions may also be specific to bulk samples rather than epitaxially-grown films, suggesting structural or chemical differences between these two classes of materials.

Hysteresis loops measured by longitudinal magneto-optical Kerr effect (LMOKE) (150 K; Fig.~\ref{fig:MagCaract}b) show in-plane magnetization reversal with a small coercive field of a few mT. The sharp transitions in the loops indicate reversal by nucleation and domain wall propagation, typical of homogeneous films. LMOKE imaging of the 12-nm film (Fig.~\ref{fig:MagCaract}c) confirms domain wall propagation during reversal and reveals large domains (10–100 $\mu$m), further indicating film homogeneity. 

\begin{table*}
    \centering
\begin{tabular}{c||c|c|c|c|c|c|c}
   $t$ & $T_C$ &$\mu_0M_s$ & $B_K$ & $g$   & $\alpha$ & $v_{DMI}$    & $D$      \\
    nm &  K    &  T        & T       &     &          & m/s          & µJ/m$^2$ \\ \hline
    3  & 258   & 0.19      & $-0.59$ & 2.3 & $<0.08$  & $18\pm4$     & $22\pm5$ \\ 
    8  & 299   & 0.36      & $-0.84$ & 2.2 & $<0.03$  & $20.5\pm1.5$ & $47\pm3$ \\ 
    12 & 297   & 0.57      & $-1.0$  & 2.1 & $<0.02$  & $10.9\pm1.3$ & $41\pm5$ \\
\end{tabular}
    \caption{
    Material parameters at 150~K for the three films. The $T_C$ and $\mu_0M_s$ were determined from SQUID magnetometry, $B_K$ and the gyromagnetic g-factor were taken from fits of the BLS frequency versus field curves, the $\alpha$ (an upper bound; see text) from the peak width, and $v_{\rm DMI}$ from fits of the Stokes/anti-Stokes frequency difference. The negative $B_K$ represents an easy-plane anisotropy. }
    \label{tab:parameters} 
\end{table*}

The samples were further characterised by studying the spin wave frequency as a function of field and wave vector using BLS spectroscopy (analysed using the laws of dispersion determined in refs.~\citenum{Kalinikos1986,Belmeguenai2015,Balan2023}; see \SI{}). The negative anisotropy field $B_K$ extracted from the BLS measurements of frequency versus field (Fig.~S2) confirms the easy-plane (i.e., ab plane) anisotropy (at 150~K; Tab.~\ref{tab:parameters}). Its magnitude is significantly larger than the expected contribution from shape anisotropy (given by $-\mu_0 M_s$), a behavior previously observed in \FGT[5] \cite{Ribeiro2022, Alahmed2021}. This suggests either the presence of an intrinsic easy-plane magnetocrystalline anisotropy or that the magnetisation is inhomogeneous across the vdW layers and gaps, leading to a stronger shape anisotropy than expected from the measured $M_s$ (see \SI).

The g-factor obtained from the BLS measurements consistently falls in the range $g=2.1 - 2.3$, showing no significant temperature dependence, in close agreement with previous reports \cite{Adhikari2024, Alahmed2021}. The notable deviation from the free-electron value of $g=2.0$ suggests a substantial orbital moment and significant spin-orbit coupling in the material. This is consistent with the orbital/spin moment ratio measured in ref.~\citenum{Ribeiro2022}. 

The Gilbert dissipation parameter ($\alpha$) can be inferred from the BLS peak width ($\Gamma$), similar to its extraction from ferromagnetic resonance measurements. However, $\Gamma$ is also broadened by inhomogeneities in the film’s magnetic properties, which are typically disentangled from the effect of $\alpha$ with  high magnetic fields taking advantage of their distinct field dependence. Since the available field was insufficient for this separation, we estimated an upper bound, $\alpha_{\rm eff}$ (see Fig.~S3 and the \SI). The lowest $\alpha_{\rm eff}$ values for each film are listed in Table~\ref{tab:parameters}. The $\alpha_{\rm eff}$ is larger for thinner films, which could be due to a higher dissipation, as is often observed in other magnetic films (e.g. ref.~\citenum{Zhao2016}). It may also be due to a higher inhomogeneity of the magnetic properties, as the properties of thinner films are more affected by variations of the thickness or of the interfaces. Overall, $\alpha_{\rm eff}$ as low as 0.02 were observed, lower than reported measurements in bulk \FGT[5]~\cite{Alahmed2021}.  

This low $\alpha$ is noteworthy, as metallic systems containing heavy elements typically exhibit large damping. This behavior is commonly interpreted using the breathing Fermi surface model~\cite{Li2019}, which predicts that $\alpha$ scales quadratically with the spin–orbit coupling (SOC) strength, linearly with the density of states (DoS) at the Fermi level, and with the electron relaxation time $\tau$ (itself  proportional to the conductivity). 
The DoS of \FGT[5] is reported to be comparable to that of iron~\cite{Joe2019}. Conversely, its low conductivity suggests a short $\tau$. In addition, its weak magnetocrystalline anisotropy points to a modest SOC. Indeed, Te primarily bonds to the weakly magnetic Fe1 sublattice \cite{Ribeiro2022}, limiting its contribution to damping. 
These factors may explain the observed $\alpha$, but they also highlight the need for an \textit{ab initio} study of damping in these materials.  Moreover, such low magnetic dissipation facilitates the motion of spin textures,  such as skyrmions, further adding to the interest of \FGT[5] as a material  for Spintronics.

\begin{figure*}[]
    \includegraphics[width=.95\textwidth]{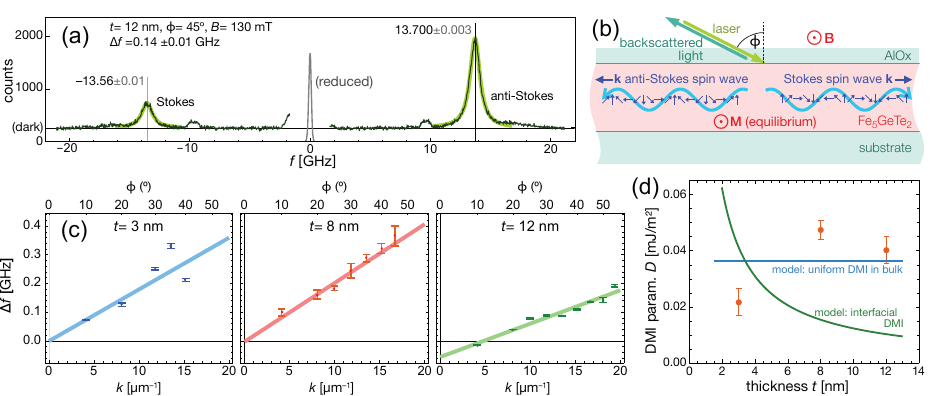}
    \caption{ 
    a) A BLS spectrum of the 12~nm film ($\phi=45$º, 150~K, $B=130$~mT). The green lines show the Lorentzian fit (fitted centres are written above the peaks). The central peak corresponds to non-scattered light and shows the instrumental width.
    b) Schematic of the BLS geometry.
    c) Stokes/anti-Stokes frequency difference, $\Delta f$, versus wave vector $k$ for the three films (points; taken at $B=$~211, 250, and 130~mT, respectively.) The linear fits (lines) are used to extract $v_{\rm DMI}$ and $D$. 
    d) Extracted $D$ versus sample thickness $t$ (points) and fits assuming a uniform bulk origin (constant blue line) and an interfacial origin ($\propto 1/t$, green line). }
    \label{fig:blsDMI} 
\end{figure*}

\subsection{DMI} 

The Stokes and anti-Stokes peaks in the BLS spectrum (Fig.~\ref{fig:blsDMI}a) correspond to cycloidal spinwaves of opposite rotation sense (Fig.~\ref{fig:blsDMI}b), and are subject to an opposite energy contributions from the DMI. In the limit of thin films ($k t \ll 1$, where $k$ is the wave vector of the spinwave determined by the incidence angle of the light), as is the case here, the difference in frequency between the Stokes and anti-Stokes peaks ($\Delta f \equiv |f_{\rm anti-Stokes}| - |f_{\rm Stokes}|$) can be fully attributed to this effect, and can be used to determine the DMI parameter $D$~\cite{Gladii2016,Belmeguenai2015}: 
\begin{equation}
    %D = \frac{M_s}{2 \gamma} \frac{\Delta f}{k}, \label{eq:vdmi}
    \Delta f = \frac{2 \gamma}{\pi M_s} D k , \label{eq:vdmi}
\end{equation}
where $\gamma = g \mu_B / \hbar $ is the gyromagnetic ratio (and  $\mu_B$ and $\hbar$ are the Bohr's magneton and the reduced Planck constant, respectively.)  As there may be a small constant $\Delta f$ due to an imperfect alignment of the spectrometer, several spectra were measured and the slope of $\Delta f(k)$ (which we dub $v_{\rm DMI}$) was determined. 
The expected reversal of the sign of $\Delta f$ with a reversal of the field direction was also verified. Fig.~\ref{fig:blsDMI}c shows the measured $\Delta f(k)$ for the 3, 8 and 12~nm films at 150~K. We observe that there is clearly a linear slope and thus a finite $D$ in all three films. All three show a slope, and thus a DMI, of the same sign. With the geometry of our setup ($\vec{k}_{\rm Stokes} \times \vec{B}$ pointing into the film surface, Fig.~\ref{fig:blsDMI}b), this positive slope corresponds to favoured right-handed spinwaves, and thus a positive DMI parameter. The calculated $D$ parameters (eq.~\ref{eq:vdmi}) are shown in Tab.~\ref{tab:parameters} for the three films.  
As a check, the DMI of the 8-nm film was also measured at 100~K and 300~K (Fig.~S4 %\ref{fig:SM fk 100 300K}
in \SI), yielding a similar value at 100 K and no detectable DMI at 300~K, as expected near $T_C$.

\section{Discussion} 
The measured DMI (0.02 to 0.05 mJ/m$^2$) is comparable to values in many spintronic systems, though lower than the strongest known effects (e.g., $\sim$0.2 mJ/m$^2$ for Co/Pt at 8 nm thickness~\cite{Belmeguenai2015}). Micromagnetic simulations\cite{Vansteenkiste2014} (see \SI) show that this DMI suffices to induce chiral magnetic textures if the anisotropy is tuned to favour out-of-plane magnetisation.

The observation of DMI that favours cycloidal textures requires broken inversion symmetry along the $c$-axis, either in the bulk of the material or due to asymmetric interfaces. This  symmetry breaking must persist over a large lateral scale, as BLS measurements  average over the 30-µm laser spot and are consistent between points several 10s of µm apart. The dependence of $D$ on film thickness $t$ can help distinguish between bulk and interfacial origins: a bulk contribution would result in a constant $D$, whereas an interfacial contribution should scale as $D \propto 1/t$ \cite{Belmeguenai2015}. Figure~\ref{fig:blsDMI}d shows $D(t)$ alongside best-fit curves for both scenarios. Clearly, $D(t)$ deviates from  the $1/t$ trend expected for interfacial DMI,  in alignment with the rather electronically similar interfaces of the films (Al$_2$O$_3$/\FGT[5] and \FGT[5]/AlO$_x$).  Instead, the $D(t)$ data is more compatible with the constant behaviour of a bulk origin, with a $D=0.04$ to 0.05 mJ/m$^2$, as measured in the thickest films. The weaker DMI in the 3-nm-thick film, which correlates with lower $M_s$ and $T_c$, can be attributed to weaker ferromagnetism in the ultrathin regime. 

Given these observations, the symmetry-breaking mechanism  extends both laterally over large scales and through the film thickness (up to 12 nm). The consistent DMI sign observed in all three films indicates a robust and reproducible effect, likely associated with the growth process, which is inherently asymmetric. We now discuss some of the possible mechanisms. The ideal structure of \FGT[5] is a rhombohedral stacking of centrosymmetric monolayers in the nonpolar ABCABC sequence (space group R$\bar{3}$m). Disruptions to this sequence by stacking faults could locally break inversion symmetry \cite{Sutter2022}. Indeed, cross-sectional STEM images reveal a high density of stacking faults in our films \cite{Ribeiro2022}, consistent with observations in bulk crystals that were slowly cooled to room temperature after growth \cite{MayPRM2019}. However, we did not identify any specific stacking sequence that would lead to a net nonzero DMI across the film thickness or a consistent DMI sign across different samples.  

Chakraborty et al. \cite{Chakraborty2022} attributed a possible bulk DMI in \FGT[3] to intercalated Fe in the vdW gaps and Fe vacancies. In our films, however, intercalated Fe was not detected \cite{Ribeiro2022} (see Fig.~S1 in \SI); it would also raise the Curie temperature~\cite{Silinskas2024}, which was not observed. Furthermore, Rutherford backscattering (RBS) analysis show a stoichiometric Fe content (Fe:Ge =  5.00±0.08)~\cite{Ribeiro2022}, indicating at most a very low concentration of Fe vacancies.

A second class of mechanisms involves symmetry breaking within the \FGT[5] vdW layers. This is supported by our RHEED observations of a $\sqrt{3} \times \sqrt{3}$ superstructure, which cannot derive from a specific vdW stacking. 
A non-centrosymmetric phase with a $\sqrt{3} \times \sqrt{3}$ superstructure was reported by Gopi et al.~\cite{Gopi2024}, who ascribed it to a pattern of 1/3 Te vacancies ordered along the perpendicular direction. Since this phase represented only a small fraction ($<10^{-2}$) of the studied bulk crystals, it was deemed to have a negligible effect on DMI. While this phase might be more prevalent in epitaxial films, RBS showed no significant Te depletion in our films~\cite{Ribeiro2022} (Te:Ge = $2.0\pm0.08$ for the 12-nm-thick film), suggesting that this mechanism is unlikely to explain our results.

\FGT[5] is expected to have two partially-filled Fe sites (“split sites”) at different vertical positions within each monolayer, “D” and “U” (Fig.~\ref{fig:Structure}a). Several groups interpreted the $\sqrt{3} \times \sqrt{3}$ superstructure as the ordering of Fe split sites \cite{Ly2021, May2019}. In the centrosymmetric structure, these sites are equally and randomly populated. However, \textit{ab initio} calculations showed that the “UDD” and “UUD” configurations are energetically more favorable \cite{Ershadrad2022}. While our RHEED measurements cannot determine whether this ordering extends beyond the top layer, electron microscopy analyses in Ref.~\citenum{May2019} showed a similar short-range order in the bulk of \FGT[5]. 
Three groups have calculated the DMI assuming a perfectly ordered “UUU” arrangement in a monolayer, using different computational methods \cite{Li2024, Ghosh2023, Gao2022}. Their predicted $|D|$ values vary significantly (0.76, 0.6, and 0.06 mJ/m$^2$), despite all employing commonly used techniques. Our measured $|D|$ (0.04 -- 0.05~mJ/m$^2$) is comparable to the lowest theoretical estimate, and much lower than the highest prediction. This quantitative difference may be due to several reasons. First, the calculations do not account for temperature, which reduces the macroscopic $D$. Second, our films are not ordered in the “UUU” configuration since this would be inconsistent with the observed $\sqrt{3} \times \sqrt{3}$ superstructure. According to Ghosh et al. \cite{Ghosh2023}, the DMI of the “UUD” or “UDD” configurations is only slightly lower than that of the “UUU” configuration ($|D|=0.05$ instead of 0.06 mJ/m$^2$), contrary to the naïve expectation of a 2/3 reduction. Finally, by symmetry, the “UUD” and “UDD” configurations contribute equally but with opposite sign DMI. While one of the two must be dominant to account for the observed DMI, the other may be present in a smaller amount and partially compensate the net DMI.

Assuming that the DMI arises from an ordering of the Fe split sites, the sign of $D$ can be used to infer the dominant split-site configuration. Both refs. \citenum{Li2024} and \citenum{Gao2022} found that a fully-ordered “UUU” configuration favours counterclockwise spinwaves, corresponding to a negative DMI. Therefore, the observed positive DMI suggests that the predominant ordering in our films is “UDD.”
\

\section{Conclusion} 
In conclusion, we have used BLS to quantitatively measure a significant DMI favouring cycloidal chiral textures in several epitaxial \FGT[5] films with nominally symmetric interfaces. The DMI values were consistent across films of different thicknesses and exhibited the same sign, indicating a bulk origin for the DMI. Although our measurements cannot reveal the microscopic origin of the DMI, they are consistent with a partial ordering of the  Fe split sites during film growth, a mechanism supported by published \textit{ab initio} calculations showing that such ordering induces DMI.

These results confirm that \FGT[5] can host a significant DMI and suggest that the reported magnetic textures in \FGT[5] are likely chiral in nature. If the mechanism for the non-centrosymmetric structure is indeed induced during the growth as we suggest, the DMI may be enhanced by adjusting the growth conditions or by employing chemical substitutions which preferentially affect the split-site sublattice, such as with Co \cite{May2020} or Ni \cite{Chen2022}. Additionally, DMI might be stronger in \FGT[5] films synthesized or processed using other techniques. The bulk DMI mechanism observed here may coexist with interfacial contributions, such as those involving oxide layers as considered in previous studies, offering opportunities to further enhance the DMI.

With its significant bulk DMI, low magnetic dissipation, and the potential to engineer its interfaces, \FGT[5] emerges as a promising material for exploring the dynamics of chiral magnetic textures and enabling spintronic applications.

\

\begin{suppinfo}

The supporting information (S.I.) includes more details and figures relating to
the experimental methods,
the structural analysis of the films,
the fitting and analysis of the BLS spectra,
additional BLS measurements, 
and micromagnetic simulations of the effects of the measured DMI.

\end{suppinfo}

\begin{acknowledgement}
    The authors thank V. Jeudy for his support  with cryogenic measurements and H. Okuno for the STEM image of Fig. S1. Magnetometry measurements were kindly performed by E. Rivière at ICMMO (CNRS) and at the Physical Measurements Platform at the LPS. 
    This work was supported by the French National Research Agency (ANR) through the projects ELMAX (ANR-20-CE24-0015), ESR/EQUIPEX+ 2D-MAG (ANR-21-ESRE0025), NEXT (ANR-23-CE09-0034), the FLAG-ERA grant MNEMOSYN (ANR-21-GRF1-0005-01), the LANEF framework for mutualised infrastructure (ANR-10-LABX-51-01), and the France 2030 government investment plan (PEPR SPIN: SPINMAT ANR-22-EXSP-0007 and CHIREX ANR-22-EXSP-0002).

\end{acknowledgement}

\bibliography{biblio} 

%%%%%%%%%%
% The main submission file shoudn't include the suppl mat. 

\newpage
\newpage
\setcounter{page}{1}

\renewcommand\thefigure{S\arabic{figure}}    
\setcounter{figure}{0}

\begin{center}
\subsection{Supporting Information for}
\section{Dzyaloshinskii-Moriya interaction in \FGT[5] epitaxial thin films}
João Sampaio  \textsuperscript{*,\textdagger}
Antoine Pascaud \textsuperscript{\textdagger},
Edgar Quero \textsuperscript{\textdagger},
André Thiaville \textsuperscript{\textdagger},
Vincent Polewczyk \textsuperscript{\textdaggerdbl},
Alain Marty \textsuperscript{\textdaggerdbl},
Frédéric Bonell \textsuperscript{\textdaggerdbl},
Alexandra Mougin \textsuperscript{\textdagger}
\end{center}

\textdagger Universit\'e Paris-Saclay, CNRS, Laboratoire de Physique des Solides, 91405 Orsay, France

\textdaggerdbl University Grenoble Alpes, CNRS, CEA, IRIG-Spintec, F-38000 Grenoble, France

* joao.sampaio@cnrs.fr

\subsection{Structural analysis}
An in-depth analysis of the structure of these films is reported in the reference \citenum{Ribeiro2022}. 

\begin{figure*}[h]
    \centering
    \includegraphics[width=.6 \textwidth]{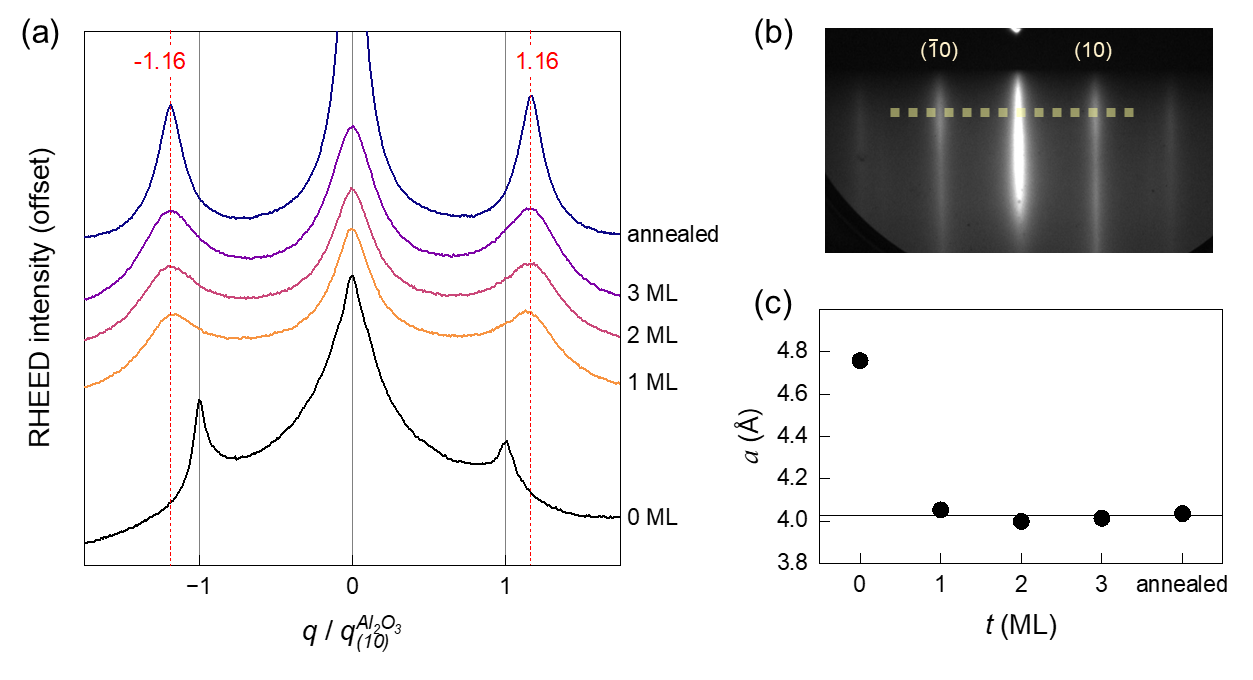}
    (d)
    \includegraphics[width=.15 \textwidth]{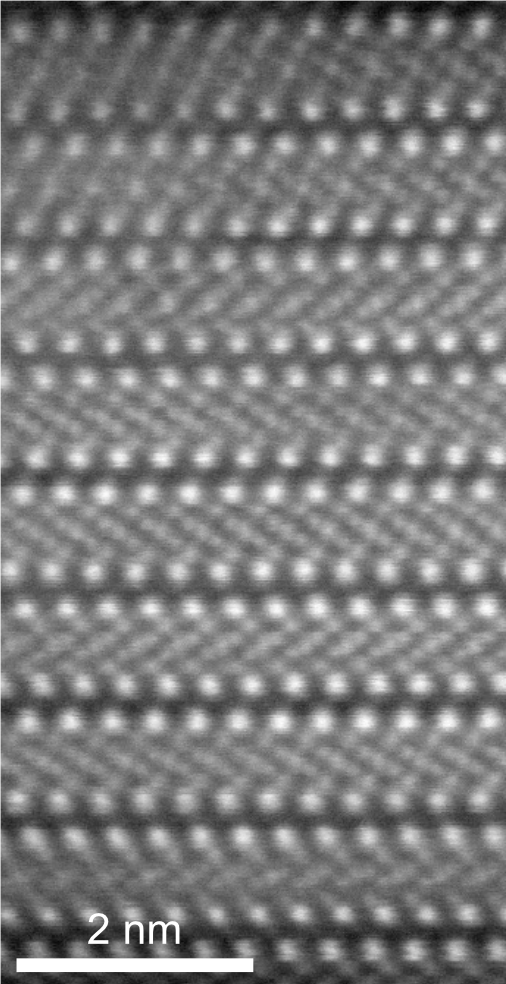}
    \caption{RHEED measurements during the growth and STEM analysis of  the \FGT[5] films. (a) Intensity profiles in the (100) direction for different thicknesses and after post-growth annealing. (b) RHEED pattern in the (100) direction illustrating the intensity profile across (10) reflections. (c) Measured in-plane lattice parameter from profiles in (a), assuming a = 4.758 Å for sapphire. The \FGT[5]  lattice parameter is fully relaxed to the bulk value from the first monolayer, indicating vdW epitaxy and negligible strain. 
    (d) High-angle annular dark-field (HAADF) cross-section image of the 12 nm-thick film in the [100] direction.}
    \label{fig:SM RHEED}
\end{figure*}

Figure \ref{fig:SM RHEED} a-c shows RHEED measurements taken during the growth process. It can be seen that the \FGT[5]  lattice parameter is fully relaxed to the bulk value from the first monolayer, indicating van der Waals epitaxy and negligible strain.

Figure \ref{fig:SM RHEED}d shows a high-angle annular dark-field (HAADF) cross-section image of the 12 nm-thick film in the [100] direction. No signs of intercalated Fe can be seen.

\subsection{Shape anisotropy of films with depth-varying magnetisation}
Assuming a homogeneous $M_s$ when it actually varies along the film thickness, $M_s(z)$, results in an underestimation of the shape anisotropy $K_d$. $K_d$ is the difference of dipolar energy density between the in-plane and perpendicular magnetisation states. The average dipolar energy density is    $-\langle \frac{\mu_0}{2}\vec{H}_d(z) \cdot \vec{M}(z) \rangle  ,$     where $\vec{H}_d(z)$ is the local dipolar field and  $ \langle \cdots \rangle$ denotes an average along the thickness ($\frac{1}{t} \int_{0}^{t} \cdots  \,dz$). In the in-plane state, $\vec{H}_d=0$, while in the perpendicular state  $\vec{H}_d(z)= -M_s(z) \unitvec{z}$. This leads to 
    $$K_d= - \frac{\mu_0}{2} \langle  M_s^2(z)\rangle .$$
    Since it is always the case that $\langle  M_s^2(z)\rangle \ge \langle M_s(z)\rangle^2$,  $K_d$ is always larger than the value calculated using the average magnetisation:  $|K_d| \ge \frac{\mu_0}{2} \langle M_s(z)\rangle^2$ (where $\langle M_s(z)\rangle$ is the magnetisation given by macroscopic magnetometry measurements.)

\subsection{Determination of magnetic parameters from BLS spectroscopy }\label{sect:BLSpars}

BLS spectroscopy was used to analyse the spinwave frequency as a function of field or wavevector. A laser beam ($\lambda=532$~nm) is directed onto the sample at an angle $\phi$ (Fig.~\ref{fig:blsDMI}b), with the sample placed inside a cryostat. 
A magnetic field is applied using permanent magnets to align the equilibrium magnetization in the film plane and perpendicular to the incident light, following the Damon-Eshbach geometry. The backscattered light is analysed with a double, three-pass Fabry-Perot spectrometer (TFP2-HC, JRS Instruments). Some of the light is inellastically backscattered by spinwaves, with a wave vector $k= 4\pi \sin \phi / \lambda$ defined by the incidence angle. The resulting spectrum exhibits a redshifted (Stokes) peak from forward-propagating spin waves and a blueshifted (anti-Stokes) peak from backward-propagating ones. The incident and analysed light are cross-polarised, which suppresses most peaks from phonon scattering. 
The confinement of the Damon–Eshbach mode to the film interfaces restricts this method to sufficiently thin samples (thinner than the laser wavelength). In this regime, BLS has become a widely used technique for the quantitative determination of DMI, as well as of other magnetic parameters.
Peak parameters — average frequency ($f_0$), frequency difference ($\Delta f$), and linewidth ($\Gamma$) — are extracted via Lorentzian fits.
The variation of these frequencies with field or wavevector allows the extraction of static ($D$, $B_K$, $M_s$, ...) and dynamic ($g$ and $\alpha$) magnetic parameters. 

\begin{figure}[ht]
    \centering
    \includegraphics[width=0.45 \textwidth]{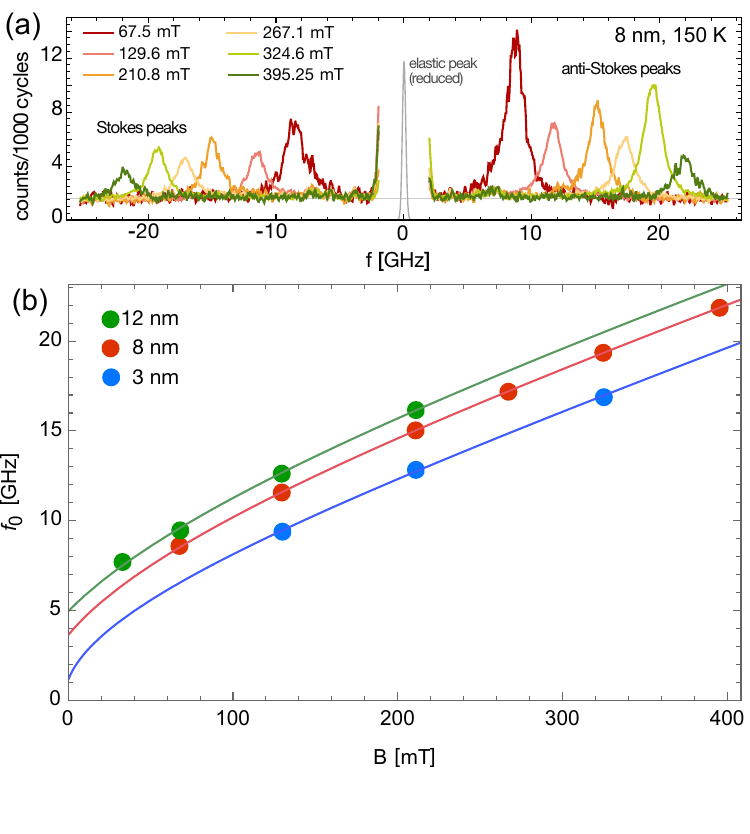}
    \caption{
    a) BLS spectra at different applied fields (8~nm film, 150~K, $\phi = 30$º). 
    The instrumental linewidth is evaluated from the elastic peak ($f=0$).
    b) Spin-wave frequency versus applied field for the three film thicknesses (150 k) ($\phi = 20$º for the 3 and 12 nm films, 30º for the 8 nm). The lines are fits using the model in ref.~\citenum{Kalinikos1986}.}
    \label{fig:BLSvB} 
\end{figure}

The effective anisotropy field $B_K$ and g-factor $g$ are extracted from $f_0(B)$ by using the law of dispersion of magnetostatic spinwaves determined by Kalinikos \& Slavin~\cite{Kalinikos1986,Belmeguenai2015}:
\begin{equation}
    f_0 =  \frac{\gamma}{2 \pi}  \sqrt{B + \mu_0 M_s P + \frac{2 A}{M_s} k^2} \sqrt{(B-B_K) - \mu_0 M_s P + \frac{2 A}{M_s} k^2},
    \label{eq:Kalinikos}
\end{equation}
where $P \approx k t /2$ for thin films ($k t \ll 1$) and $A$ is the exchange stiffness (neglected in our analysis).
Fig.~\ref{fig:BLSvB}a shows BLS spectra of the 8-nm sample taken at different $B$. All spectra clearly show the Stokes and anti-Stokes peaks, which are of different amplitude, typical of magnon BLS (but not of phonons). 
$B_K$ and $g$ parameters were extracted from fits of $f_0(B)$ (lines in Fig.~\ref{fig:BLSvB}b) and are shown in Tab.~\ref{tab:parameters} for all three films at 150~K.

\subsection{Determination of the Gilbert dissipation parameter}\label{sect:alpha}
The Gilbert dissipation parameter $\alpha$ can be obtained from the peak width $\Gamma$. The main contributions to $\Gamma$ are the instrumental resolution ($\Gamma_0$), the magnetic dissipation, and the inhomogeneity of the magnetic properties such as the anisotropy ($\delta B$): 
\begin{equation}
    \Gamma= \Gamma_0 +  \frac{\gamma}{2\pi } \alpha (2 B - B_K) + \gamma \delta B \ F(B,B_K),
    \label{eq:gamma}
\end{equation}
where $F()$ is a function that tends towards a constant for $B \gg B_K$ (see SI of ref.~\citenum{Balan2023}; a $2\pi$ factor is missing in the definitions of $f_x$ and $f_z$ in that reference.) 

$\Gamma_0$ is measured from the elastic peak at $f=0$ (seen in Fig.~\ref{fig:BLSvB}a). $\delta B$ may be disentangled from the dissipation by applying a large field such that $\alpha (2 B - B_K) \gg \delta B \ F(B,B_K)$, however this was not possible with our setup. An upper bound of the dissipation parameter, $\alpha_{\rm eff}$, can be determined by neglecting $\delta B$:
\begin{equation}
    \alpha_{\rm eff} = \frac{2\pi}{\gamma} \frac{\Gamma-\Gamma_0}{2 B - B_K}
    % (\Gamma-\Gamma_0)/(\gamma (2 B - B_K)) \ge \alpha
\end{equation}

\begin{figure}[ht]
    \centering
    \includegraphics[width=0.5\textwidth]{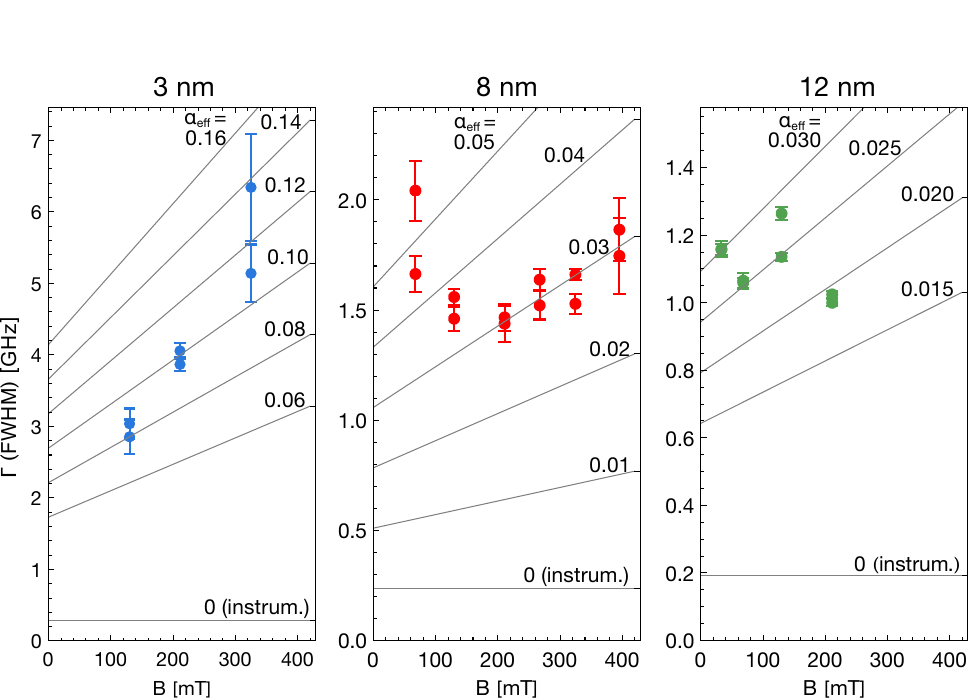}
    \caption{ Peak FWHM $\Gamma$ versus field $B$ for the three films. The grey lines are the expected variation assuming different values of $\alpha_{\rm eff}$ (see text for details on the calculation of $\alpha_{\rm eff}$).}
    \label{fig:alpha}
\end{figure}

The measured $\Gamma$ vs. $B$, along with the calculated $\alpha_{\rm eff}$, is shown in 
Fig.~\ref{fig:alpha}, and the minimum values of $\alpha_{\rm eff}$ are included in Tab.~\ref{tab:parameters}.

\subsection{Additional BLS measurements at different temperatures}

\begin{figure*}[ht]
    \centering
    \includegraphics[width=0.55 \textwidth]{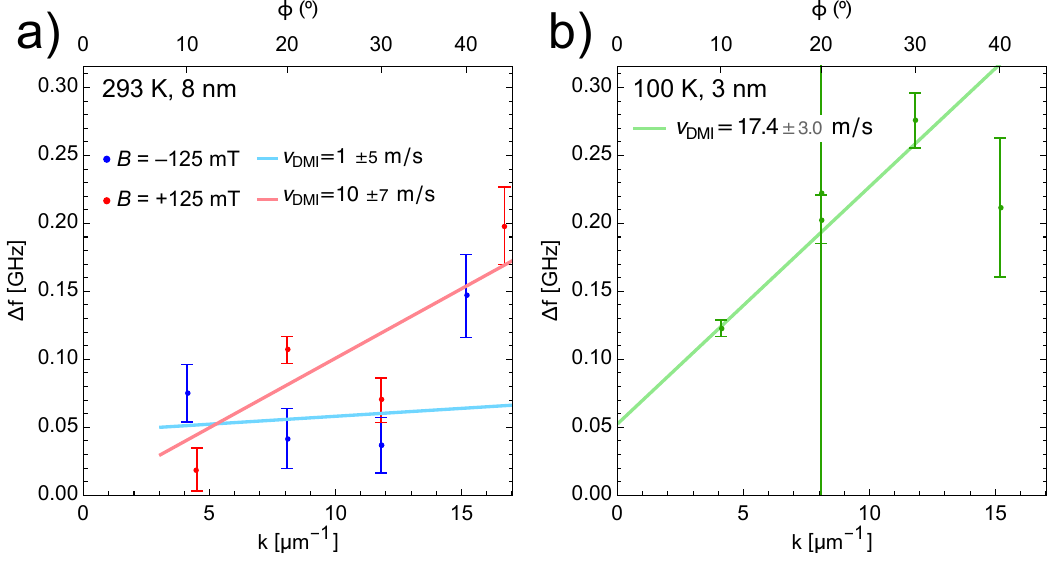}
    \caption{Stokes/anti-Stokes frequency difference, $\Delta f$, versus wave vector $k$ for the 8~nm film at 100 and 300~K, showing a finite slope (and thus DMI) at 100~K. At 300~K, the slope is negligible (the fit p-values, i.e. the probability of a zero slope, is 0.85 and 0.29). }
    \label{fig:SM fk 100 300K}
\end{figure*}

Some additional measurements of $f(k)$ were performed at 100~K and near $T_C$ (300~K) for the 8~nm film, and the Stokes/anti-Stokes frequency difference $\Delta f(k)$ is shown in Fig.~\ref{fig:SM fk 100 300K}. 
At 100~K, a finite $v_{\rm DMI}=\Delta f/k$ of similar magnitude to that at 150~K is observed ($17\pm3$~m/s compared to $20.5\pm1.5$~m/s).
At 300~K, two series of $f(k)$ with opposite applied field were measured. If there were DMI, the two series should present slopes of opposite sign. Instead, the fits are compatible with no slope ($v_{\rm DMI}= 1\pm5$ and $10\pm 7$ m/s, with the statistical tests of the fits giving a high probability for no slope). 
This is expected, as, like the conventional exchange interaction, the DMI energy should scale with $M_S^2$ (and $v_{\rm DMI}$ with $M_S$), which drops sharply near $T_C$ but varies little between 100~K and 150~K.

\subsection{Micromagnetic simulations of stabilised chiral textures}

\begin{figure*}[ht]
    \centering
    \includegraphics[width= .7 \textwidth]{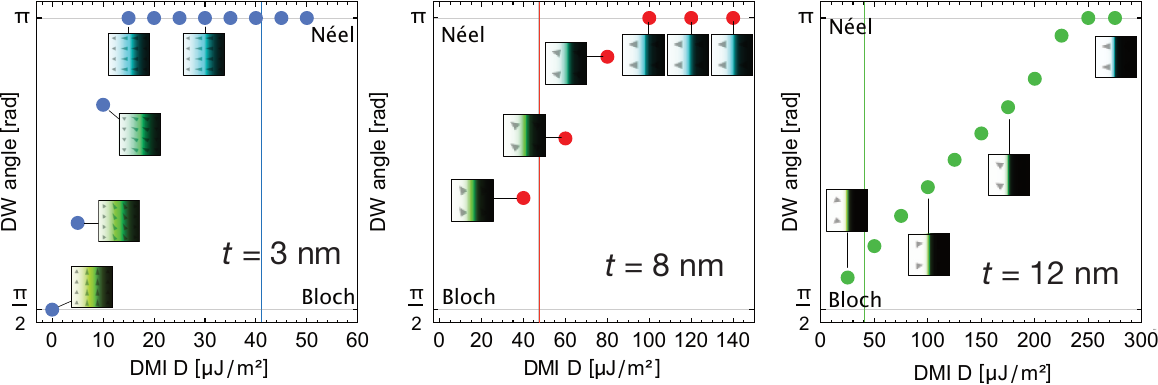}
    \caption{Micromagnetic simulations of the domain wall core angle as a function of DMI strength $D$, showing the continuous transition from Bloch to Néel walls. Simulations assume an effective perpendicular anisotropy of 200 J/m$^3$ and exchange stiffness $A=2$~pJ/m. Insets: examples of DW profiles. Vertical lines indicate experimentally measured values of D.}
    \label{fig:SM sim DW}
\end{figure*}

\begin{figure*}
    \centering
    \includegraphics[width=0.7 \textwidth]{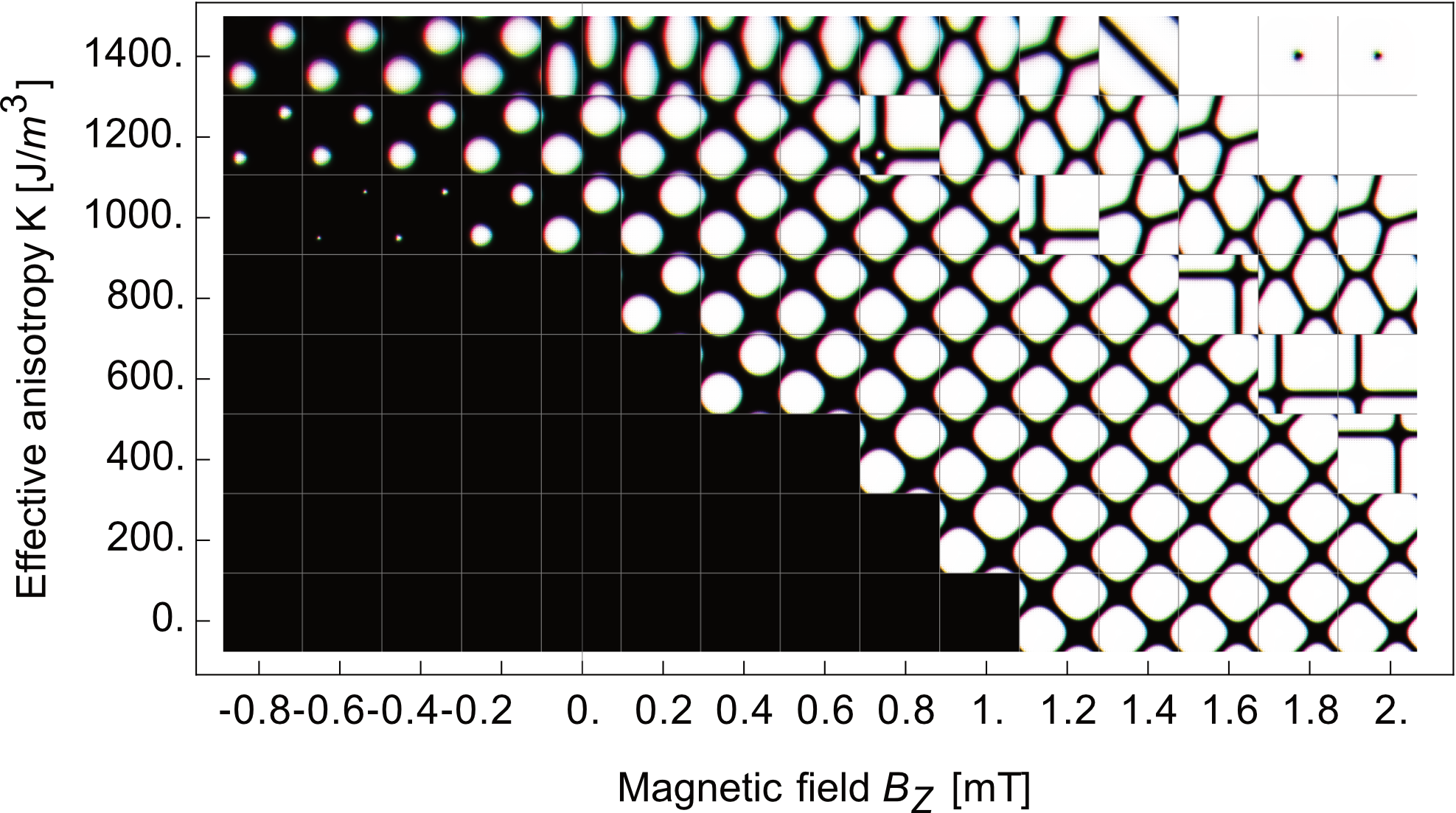}
    \caption{
    Simulated stable spin textures for the 3-nm-thick film as a function of the induced perpendicular anisotropy $K$ and applied magnetic field. Simulations were run in a $2.05 \times 2.05$ $\mu m^2$ square, starting from two positive-core skyrmions (white) in a uniform background $m = -z$ (black). Isolated skyrmions are stabilised for $K>600 \, J/m^3$; at lower $K$, the stable skyrmion diameter likely exceeds the simulation area.}
    \label{fig:SM sim Sk}
\end{figure*}

To evaluate whether the observed DMI magnitude is sufficient to stabilise chiral spin textures, we performed micromagnetic simulations \cite{Vansteenkiste2014} of the domain wall  structure for varying values of $D$, assuming a small induced perpendicular anisotropy (200 J/m$^3$) for the parameters of the three films. Such a modest anisotropy could be through interfacial effects (a common approach in thin-film systems) or by electric field bias. 
The simulations, shown in Figure~\ref{fig:SM sim DW}, reproduce the behaviour described by Thiaville et al.~\cite{Thiaville2012}: a continuous transition from Bloch to chiral Néel walls as D increases, up to a saturation value. In the simulations, the 3-nm-thick film exhibits fully Néel-type DWs, while the thicker films display hybrid chiral Bloch–Néel configurations. Based on the parameters of the 3-nm film, we also simulated a skyrmion lattice state  under various values of induced perpendicular anisotropy and applied bias magnetic field, Fig. \ref{fig:SM sim Sk}. We found that small isolated skyrmions can be stabilised for higher effective anisotropy (> 1 kJ/m³), while larger skyrmions appear for lower anisotropy values.

\end{document}